\input harvmac

\lref\HenneauxTeitelboim{
M. Henneaux and C. Teitelboim, Commun. Math.
Phys. {\bf 98} (1985) 391.}

\lref\Dirac{
P.A.M. Dirac, J. Math. Phys. {\bf 4} (1963) 901.}

\lref\Fronsdal{
C. Fronsdal, Phys. Rev. {\bf D12} (1975) 3819.}

\lref\oldcft{
M. Flato and C. Fronsdal, J. Math. Phys. {\bf 22} (1981) 1100;
E. Angelopoulos, M. Flato, C. Fronsdal and D. Sternheimer,
Phys. Rev. {\bf D23} (1981) 1278.}

\lref\singbrane{
C. Fronsdal, Phys. Rev. {\bf D26} (1982) 1988;
H. Nicolai and E. Sezgin, Phys. Lett. {\bf B143} (1984) 389;
M. G\"{u}naydin and N. Marcus, Class. Quant. Grav. {\bf 2}
(1985) L1;  M.~G\"{u}naydin, P. van Nieuwenhuizen and N.P. Warner,
Nucl. Phys. {\bf B255} (1985) 63.}

\lref\mtwo{
E. Bergshoeff, M.J. Duff, C.N. Pope and E. Sezgin,
Phys. Lett. {\bf B199} (1987) 69; \hfill\break
M.P. Blencowe and M.J. Duff,
Phys. Lett. {\bf B203} (1988) 229; E. Bergshoeff, \hfill\break
A. Salam, E. Sezgin
and Y. Tanii, Phys. Lett. {\bf B205} (1988) 237 and
Nucl. Phys. {\bf B305} [{\bf FS23}] (1988) 497.}

\lref\NST{
H. Nicolai, E. Sezgin and Y. Tanii, Nucl. Phys. {\bf B305}
[{\bf FS23}] (1988) 483.}

\lref\mfive{
G.W. Gibbons and P.K. Townsend, Phys. Rev. Lett.
{\bf 71} (1993) 3754, {\tt hep-th/9307049}.}

\lref\fone{
M. G\"{u}naydin, B.E.W. Nilsson, G. Sierra and P.K. Townsend,
Phys. Lett. {\bf B176} (1986) 45.}

\lref\SfSk{
K. Sfetsos and K. Skenderis, ``Microscopic derivation of the
Bekenstein-Hawking
entropy formula for non-extremal black holes'', Nucl. Phys. {\bf B}
in press,
{\tt hep-th/9711138}.
}

\lref\maldacena{
J. Maldacena, ``The large $N$ limit of superconformal
field theories and supergravity'', {\tt hep-th/9711200}.}

\lref\browhen{
J.D. Brown and M. Henneaux,
Commun. Math. Phys. {\bf 104} (1986) 207;
O. Coussaert, M.~Henneaux and P. van Driel, Class. Quant. Grav. {\bf
12}
(1995) 2961, {\tt qr-qc/9506019}.}

\lref\GubserKlebanovPolyakov{
S.S. Gubser, I.R. Klebanov and A.M. Polyakov, ``Gauge
theory correlators from non-critical string theory'', {\tt
hep-th/9802109}.}

\lref\WittenI{
E. Witten, ``Anti-de Sitter space and holography'', {\tt
hep-th/9802150}.}

\lref\FerFro{
S. Ferrara and C. Fronsdal, ``Gauge fields as composite boundary
excitations'',
{\tt hep-th/9802126}.}

\lref\velgoi{
P. Claus, R. Kallosh and A. Van Proeyen,
``M 5-brane and superconformal (0,2) tensor multiplet in 6
dimensions'',
{\tt hep-th/9711161};
P. Claus, R. Kallosh, J. Kumar, P. Townsend and A.Van Proeyen,
``Conformal theory of M2, D3, M5 and D1+D5 branes'', {\tt
hep-th/9801206}.}

\lref\ferrara{
S. Ferrara and C. Fronsdal, ``Conformal Maxwell theory as
a singleton field theory on $AdS_5$, IIB three-branes and duality'',
{\tt hep-th/9712239};
M. G\"unaydin and D. Minic, 
``Singletons, Doubletons and M-theory'', {\tt hep-th/9802047};
S. Ferrara, C. Fronsdal and A. Zaffaroni, ``On N=8 Supergravity on
$AdS_5$ and N=4 Superconformal Yang-Mills theory'', {\tt
hep-th/9802203};
S. Ferrara, A. Kehagias, H. Partouche and A. Zaffaroni,
``Membranes and Fivebranes with Lower Supersymmetry and their $AdS$
Supergravity Duals'', {\tt hep-th/9803109};
L. Andrianopoli and S. Ferrara, ``K-K excitations
on $AdS_5 \times S^5$
as $N=4$ ``primary'' superfields'', {\tt hep-th/9803171}.}

\lref\kaSil{
S. Kachru and E. Silverstein, ``4d Conformal
Field Theories and Strings on Orbifolds'', {\tt hep-th/9802183};
M. Berkooz, ``Supergravity
Dual of a (1,0) Field Theory in Six Dimensions'', {\tt
hep-th/9802195};
A. Lawrence, N. Nekrasov and C. Vafa, ``On Conformal Theories
 in Four Dimensions'', {\tt hep-th/9803015};
M. Bershadsky, Z. Kakushadze and C. Vafa, ``String
Expansion as Large N Expansion of Gauge Theories'', {\tt
hep-th/9803076};
J. Gomis, ``Anti-de Sitter Geometry and Strongly Coupled Gauge
Theories'',
{\tt hep-th/9803119};
Y. Oz and J. Terning,
``Orbifolds of $AdS_5 \times S^5$ and $4d$ Conformal Field
Theories'',
{\tt hep-th/9803167};
M.J. Duff, H. L\"u and C.N. Pope, {\it $AdS_5 \times S^5$
 untwisted}, {\tt hep-th/9803061}; E. Halyo, ``Supergravity on
$AdS_{5/4} \times$ Hopf Fibrations and Conformal Field Theories'',
{\tt hep-th/9803193};
H. J. Boonstra, B. Peeters and K. Skenderis, ``Brane intersections,
Anti-de Sitter spacetimes and dual superconformal theories'', 
{\tt hep-th/9803231}.}

\lref\rossoi{
I.Ya. Aref'eva and I.V. Volovich, ``On Large N Conformal
Theories, Field Theories in Anti-De Sitter Space and Singletons'',
{\tt hep-th/9803028}; I.V. Volovich, ``Large N Gauge Theories and
Anti-de Sitter Bag Model'', {\tt hep-th/9803174}.}

\lref\AharOz{
O. Aharony, Y. Oz and Z. Yin, ``M Theory
on $AdS_p \times S^{11-p}$ and Superconformal Field Theories'',
{\tt hep-th/9803051}; S. Minwalla, ``Particles on $AdS_{4/7}$
and Primary Operators on $M_{2/5}$ Brane Worldvolumes'',
{\tt hep-th/9803053};
E. Halyo, ``Supergravity on $AdS_{4/7} \times S^{7/4}$ and M
Branes'',
{\tt hep-th/9803077}.}

\lref\wilson{
S.-J. Rey and J.-T. Yee, ``Macroscopic Strings as Heavy Quarks of
Large N Gauge Theory and Anti-de Sitter Supergravity'',
{\tt hep-th/9803001};
J. Maldacena, ``Wilson loops in large N field theories'',
{\tt hep-th/9803002};
J. Minahan, ``Quark-Monopole Potentials in Large N Super
Yang-Mills'',
{\tt hep-th/9803111};
S.-J. Rey, S. Theisen and J.-T. Yee, ``Wilson-Polyakov Loop at
Finite Temperature in Large N Gauge Theory and Anti-de Sitter
Supergravity'',
{\tt hep-th/9803135};
A. Brandhuber, N. Itzhaki, J. Sonnenschein and S. Yankielowicz,
``Wilson Loops in the Large N Limit at Finite Temperature'',
{\tt hep-th/9803137};
A. Volovich, ``Near anti-de Sitter geometry and corrections to
the large $N$ Wilson loop'', {\tt hep-th/9803220}.}

\lref\wittenII{
E. Witten, ``Anti-de Sitter Space, Thermal Phase Transition,
And Confinement in Gauge Theories'', {\tt hep-th/9803131}.}

\lref\FerraraGattoGrilloParisi{
S. Ferrara, R. Gatto, A.F. Grillo and G. Parisi,
Lett. Nuovo Cim. {\bf 4} (1972) 115.}

\lref\LeighRozali{
R. Leigh and M. Rozali,
``The large $N$ limit of the $(2,0)$ superconformal field theory'',
{\tt hep-th/9803068}.}

\lref\MackSalam{
G. Mack and A. Salam,
Ann. Phys. (N. Y.) {\bf 53} (1969) 174.}


\def\Dsl{D \!\!\!\! /}
\def\G{{\bf \Gamma}}
\def\x{{\bf x}}
\def\xp{{\bf x}^\prime}
\def\xpp{{\bf x}^{\prime \prime}}

\rightline{\vbox{\hbox{CERN-TH/98-78}\hbox{\tt hep-th/9803251}}}
\Title{}{Spinors and the AdS/CFT correspondence}
\bigskip
\centerline{{\bf M{\aa}ns Henningson} and {\bf Konstadinos Sfetsos}}
\bigskip
{\it \centerline{Theory Division, CERN}
\centerline{CH-1211 Geneva 23, Switzerland} \centerline{\tt
henning@nxth04.cern.ch, sfetsos@mail.cern.ch}}

\bigskip \bigskip \bigskip \centerline{\bf Abstract}
We consider a free massive spinor field in Euclidean Anti-de Sitter
space.
The usual Dirac action in bulk is supplemented by a certain boundary
term.
The boundary conditions of the field are parametrized by a spinor on
the
boundary, subject to a projection.
We calculate the dependence of the partition function
on this boundary spinor. The result agrees with the generating
functional of the correlation functions of a quasi-primary spinor
operator, of
a certain scaling dimension, in a free conformal field theory on the
boundary.

\Date{\vbox{\line{CERN-TH/98-78\hfill} \line{March 1998 \hfill}}}

\newsec{Introduction}

The relationship between $d$-dimensional conformal field theories and
field theories in $(d + 1)$-dimensional Anti-de Sitter space
has been the focus of studies from various viewpoints over the years.
Conformal field theories provide a field theoretical interpretation
for the
so-called ``singleton'' representations \refs{\Dirac,\Fronsdal}
of the symmetry group $SO(d,2)$ \oldcft, which correspond to gauge
degrees
of freedom everywhere except in the boundary \singbrane.
A further development of this correspondence was that the
singletons of AdS-spaces were related to brane solutions of
supergravity via the
indentification of the various brane world-volume fields with
members of the corresponding (super)-singleton
multiplets \refs{\mtwo,\NST,\mfive,\fone}.
The AdS/CFT
correspondence was exploited in \SfSk\ in connection with the
physics of non-extremal black holes.
In parallel, it has been conjectured that there is a relation between
the large-$N$ limit of certain superconformal gauge theories 
(realized by branes) and the
supergravity limit of $M$-theory or string theory \maldacena.

A more precise AdS/CFT relationship was suggested in
\refs{\GubserKlebanovPolyakov,
\FerFro,\WittenI} as follows\foot{
The case of $AdS_3$ is quite well understood \browhen.
Additional recent work concerning various
aspects of the AdS/CFT correspondence can be found
in
\refs{\velgoi,\ferrara,\kaSil,\AharOz,\wittenII,\wilson,\rossoi}.}:
$AdS_{d + 1}$ has
a boundary $M_d$ at spatial infinity.
This means that the action functional $S[\phi]$ of a field theory on
$AdS_{d + 1}$ must be supplemented by some boundary conditions on the
field
$\phi$.
(For notational simplicity we consider only the case of a single
field.
The generalization to several fields is straightforward.)
To specify a boundary condition, we first choose a function $x^0$
on $AdS_{d + 1}$ with a simple zero on the boundary.
The boundary condition then amounts to
\eqn\bc{
\lim_{x^0 \rightarrow 0} (x^0)^{\Delta - d} \phi = \phi_0 \ ,
}
for some (finite) field $\phi_0$ defined on the boundary $M_d$. The
value of the constant $\Delta$ is fixed by the requirement that the
classical equations of motion allow this behaviour of $\phi$. We can
now calculate the partition function
\eqn\partftn{
Z_{AdS} [\phi_0] = \int_{\phi_0} {\cal D}  \phi \exp (-S[\phi])\ ,
}
where the subscript on the integral indicates that we should only
integrate over field configurations that fulfil the boundary
condition \bc.

We can obtain a finite metric $g$ on $M_d$ by multiplying the metric
$G$ on $AdS_{d + 1}$ by $(x^0)^2$ and restricting to $M_d$. The
freedom to choose the function $x^0$, subject to the restriction that
it has a simple zero on $M_d$, means that only the conformal
structure of $M_d$ is uniquely determined. The $O(d + 1, 1)$ isometry
group of $AdS_{d + 1}$ acts as the conformal group on $M_d$. We can
therefore equate the partition function $Z_{AdS} [\phi_0]$ with the
generating functional
\eqn\CFTgen{
Z_{CFT} [\phi_0] = {\bigl\langle} \exp \int_{M_d} d^d \x \sqrt{g} {\cal O}
\phi_0 {\bigl\rangle}\ ,
}
of a quasi-primary operator ${\cal O}$ in some conformal field theory
on $M_d$. Such an operator is characterized by its quantum numbers
under an $O(d) \times SO(1, 1)$ subgroup of the conformal group, i.e.
by its Lorentz group representation and scaling dimension \MackSalam.
Invariance of the $\int d^d \x \sqrt{g} {\cal O} \phi_0$ coupling
implies
that ${\cal O}$ and $\phi_0$ transform in conjugate Lorentz
representations
and that the sum of their scaling dimensions is $d$. The boundary
behaviour \bc\ and the fact that the metric on $AdS_{d + 1}$ has
a double pole on $M_d$ then determines the scaling dimension of
${\cal O}$
to be equal to $\Delta$.

Some examples of this AdS/CFT correspondence were given in \WittenI,
where,
in particular, the cases of a free massive scalar field and of a free
$U(1)$ gauge theory on $AdS_{d + 1}$ were considered.
The exact partition function \partftn\ is then given by the tree-level contribution, i.e. by the exponential of the action evaluated
for a field configuration that solves the classical equations of
motion with boundary conditions given by \bc. For such a field
configuration, the action can in fact be rewritten as a total
derivative and thus reduces to a boundary term. The result is that
the partition function \partftn\ indeed equals the generating
functional \CFTgen\ of the correlation functions of a free
quasi-primary operator of a certain scaling dimension.

Fermionic fields are obviously of equal importance to scalar and
gauge fields.
Hence, it will be the purpose of this paper to perform an analysis
for the
free Dirac spinor field $\psi$ and its conjugate $\bar{\psi}$ of mass
$m$ on
$AdS_{d + 1}$.
It turns out that the usual
Dirac action in bulk must be supplemented by a certain boundary term.
Also, one would expect that the boundary conditions are determined by
a spinor field $\psi_0$ and its conjugate $\bar{\psi}_0$ on $M_d$.
For $d$ even, $\psi_0$ would be a Dirac spinor, whereas for $d$ odd,
it would be a pair of Dirac spinors. Actually, it will turn out that
half of $\psi_0$
(and of $\bar{\psi}_0$) will have to be put to zero, so we are left
with a chiral spinor or a single Dirac spinor for $d$ even or odd
respectively. Which half is retained is determined by the sign of the
mass $m$. In this way we
find that the partition function indeed reproduces the correlation
functions of a free quasi-primary spinor operator ${\cal O}$ of
scaling dimension $\Delta = {d \over 2} + |m|$.

For some of these issues there is an overlap with 
a recent paper \LeighRozali,
which appeared while we were completing this work.

\newsec{The action}
Our starting point is the usual free spinor action on $AdS_{d + 1}$
\eqn\Szero{
S_0 = \int_{AdS} d^{d + 1} x \sqrt{G} \bar{\psi} (\Dsl - m) \psi\ ,
}
from which follow the Dirac equations of motion
\eqn\eom{
\eqalign{
(\Dsl - m) \psi & = 0 \ ,\cr
\bar{\psi} (- \overleftarrow{\Dsl} - m) & = 0\ . \cr
}
}
Before we determine the boundary conditions to be imposed at
infinity, we will discuss a difficulty with the action \Szero. Since
this is a free field theory, the exact partition function is given by
the tree-level contribution. Up to a constant, that arises from the
Gaussian path integral over the square integrable fluctuations of the
fields, this equals the exponential of the action functional
evaluated for a field configuration that solves the classical
equations of motion and obeys the boundary conditions. However, for a
spinor field, the action \Szero\ vanishes for any field configuration
that satisfies the equations of motion \eom. The same is true for the
total derivative term
\eqn\totalderiv{
\int_{AdS} d^{d + 1} x \sqrt{G} \left( \bar{\psi}
\overleftarrow{\Dsl} \psi + \bar{\psi} \Dsl \psi \right) ,
}
a multiple of which could be added to the action without changing the
equations of motion. It would therefore seem that the partition
function is actually independent of the boundary conditions and would
not reproduce any conformal field theory correlation functions.

To avoid this conclusion, we propose that the action \Szero\ be
supplemented by a multiple of
\eqn\Sone{
S_1 = \lim_{\epsilon \rightarrow 0} \int_{M_d^\epsilon} d^d \x
\sqrt{G_\epsilon} \bar{\psi} \psi\ ,
}
where $M_d^\epsilon$ is a closed $d$-dimensional submanifold of
$AdS_{d + 1}$, which approaches the boundary manifold $M_d$ of $AdS_{d
+ 1}$ as $\epsilon$ goes to zero. The metric $G_\epsilon$ on
$M_d^\epsilon$ appearing in the expression \Sone\ is the one induced
from the metric $G$ of $AdS_{d + 1}$. While perhaps slightly
unfamiliar, the addition of this term to the action preserves the
crucial properties of the original action \Szero: it is invariant
under the $O(d + 1, 1)$ isometry group of $AdS_{d + 1}$, since this
group maps $M_d$ to itself. The equations of motion in the bulk of
$AdS_{d + 1}$ are still given by \eom. While the term \Sone\ clearly
depends on the boundary conditions on the fields, it does not affect
the path integral over square integrable quantum fluctuations. In the
present context, the relative coefficient between the terms \Szero\
and \Sone\ is not determined, and we will only assume that it is
non-zero. In other theories, gauge invariance or supersymmetry, for
example, will impose restrictions on the coefficients of various
boundary terms.

\newsec{The computation}
In this section, we will show that the theory on $AdS_{d + 1}$
described above is indeed equivalent to the free conformal field
theory on $M_d$ of a quasi-primary spinor operator ${\cal O}$.

We choose coordinates $x^\mu = (x^0 , \x) = (x^0, x^i)$, $i = 1,
\ldots, d$, such that Euclidean $AdS_{d + 1}$ is represented by the
domain $x^0 > 0$ and the metric $d s^2$ is given by
\eqn\Gmetric{
d s^2 = G_{\mu \nu} d x^\mu d x^\nu = (x^0)^{-2} \left( d x^0 d x^0 +
g_{i j} d x^i d x^j \right) = (x^0)^{-2} \left( d x^0 d x^0 + d \x
\cdot d \x \right)\ .
}
The boundary $M_d$ is defined by the hypersurface $x^0 = 0$ plus a
single point at $x^0 = \infty$. The metric $d \tilde{s}^2$ on $M_d$
is obtained by multiplying $d s^2$ by $(x^0)^2$ and restricting to
$M_d$ so that
\eqn\gmetric{
d \tilde{s}^2 = g_{i j} d x^i d x^j = d \x \cdot d \x\ .
}

To couple a spinor field to this background, we need to choose a
local Lorentz frame, i.e. a vielbein $e^a_\mu$, $a = 0, \ldots, d$
such that $G_{\mu \nu} = e^a_\mu e^b_\nu \eta_{a b}$. A convenient
choice is
\eqn\vielbein{
e_\mu^a = (x^0)^{-1} \delta_\mu^a\ ,
}
for which the corresponding spin connection $\omega_\mu^{a b}$ has
\eqn\spinconn{
\omega_i^{0 j} = - \omega_i^{j 0} = (x^0)^{-1} \delta_i^j
}
and all other components vanishing.
With these choices, the operator $\Dsl$ is given by
\eqn\Dslash{
\Dsl = e_a^\mu \Gamma^a \left( \partial_\mu + {1 \over 2}
\omega_\mu^{b c} \Sigma_{b c} \right) = x^0 \Gamma^0 \partial_0 + x^0
\G \cdot {\bf \nabla} - {d \over 2} \Gamma^0\ ,
}
where $\Gamma^a = (\Gamma^0, \Gamma^i) = (\Gamma^0 , \G)$ fulfil $\{
\Gamma^a, \Gamma^b \} = 2 \eta^{a b}$ and $\partial_\mu =
(\partial_0, \partial_i) = (\partial_0, {\bf \nabla})$.

Next, we will construct solutions to the equations of motion \eom\
with various
boundary behaviours. This can be accomplished by first constructing a
solution with singular behaviour at a single point on the boundary
$M_d$ (most conveniently the point at $x^0 = \infty$), and then
applying an element of the $O(d + 1, 1)$ isometry group of $AdS_{d +
1}$ to move the singularity to an arbitrary point on the boundary.
(This transformation has to be accompanied by a compensating local
Lorentz transformation to preserve the gauge choice \vielbein.) In
this way one constructs the field configurations
\eqn\configurations{
\eqalign{
\psi (x^0, \x) & = \int d^d \xp \Bigl(x^0 \Gamma^0 + (\x \!-\! \xp)
\cdot \G \Bigr) \Bigl((x^0)^2 + |\x \!-\! \xp|^2 \Bigr)^{-{d + 1
\over 2} + m \Gamma^0} (x^0)^{{d \over 2} - m \Gamma^0} \psi_0 (\xp)\
, \cr
\bar{\psi} (x^0, \x) & = \int d^d \xpp \bar{\psi}_0 (\xpp) (x^0)^{{d
\over 2} + m \Gamma^0} \Bigl((x^0)^2 + |\x \!-\! \xpp|^2\Bigr)^{-{d +
1 \over 2} - m \Gamma^0} \Bigl(x^0 \Gamma^0 + (\x \!-\! \xpp) \cdot
\G\Bigr)\ , \cr
}
}
which solve \eom\ for arbitrary finite spinors
$\psi_0 (\xp)$ and $\bar{\psi}_0 (\xpp)$.

To determine the boundary behaviour of these solutions, it is
convenient to decompose $\psi_0 (\xp)$ and $\bar{\psi}_0 (\xpp)$ as
\eqn\decomp{
\eqalign{
\psi_0 (\xp) & = \psi_+ (\xp) + \psi_- (\xp)\ , \cr
\bar{\psi_0} (\xpp) & = \bar{\psi}_+ (\xpp) + \bar{\psi}_- (\xpp)\ ,
\cr
}
}
where $\Gamma^0 \psi_\pm (\xp) = \pm \psi_\pm (\xp)$ and
$\bar{\psi}_\pm (\xpp) \Gamma^0 = \pm \bar{\psi}_\pm (\xpp)$. Without
loss of generality we take the mass $m$ to be positive. One then
finds that
\eqn\boundbeh{
\eqalign{
\lim_{x^0 \rightarrow 0} (x^0)^{- {d \over 2} + m} \psi (x^0, \x) & =
-c\
 \psi_- (\x) + \int d^d \xp |\x \! -\! \xp|^{- d - 1 + 2 m} (\x\! -\!
\xp) \cdot \G \psi_+ (\xp)\ , \cr
\lim_{x^0 \rightarrow 0} (x^0)^{- {d \over 2} + m} \bar{\psi} (x^0,
\x) & = c
\ \bar{\psi}_+ (\x) + \int d^d \xpp \bar{\psi}_- (\xpp )(\x\! -\!
\xpp) \cdot \G
|\x \! -\! \xpp|^{- d - 1 + 2 m}\ , \cr
}
}
where the
constant $c= \pi^{d/2} {\Gamma(m+\half)/ \Gamma(m+{d+1\over 2})}$.
For the right-hand side of \boundbeh\ to be square integrable,
with respect to the measure $d^d \x$ on $M_d$, we will furthermore
have to impose the conditions
\eqn\zerocond{
\eqalign{
\psi_+ (\xp) & = 0\ , \cr
\bar{\psi}_- (\xpp) & = 0 \ . \cr
}
}
(A similar condition has appeared in \HenneauxTeitelboim\ for the case of
a Rarita-Schwinger field.) Before \zerocond\ are imposed, $\psi_0 (\xp)$ and $\bar{\psi}_0 (\xp)$
both transform as a Dirac spinor or two Dirac spinors when $d$ is
even or odd, respectively. The two irreducible terms of this
representation are distinguished by their $\Gamma^0$ eigenvalues.
(When $d$ is even, $\Gamma^0$ is simply the chirality operator on
$M_d$.) The conditions \zerocond\ thus mean that we are left with a
chiral spinor when $d$ is even and a single Dirac spinor when $d$ is
odd.

As discussed in the previous section, the bulk action \Szero\
vanishes for such a configuration, so the entire contribution to the
partition function comes from the boundary term \Sone.
To calculate this term, we take the hypersurface $M^\epsilon_d$ to be
given by
$x^0=\epsilon = {\rm constant}$.
The induced metric on $M^\epsilon_d$ is then
$ds^2_\epsilon =\epsilon^{-2} d\x\cdot d \x$, with determinant
$G_\epsilon= \epsilon^{-2 d}$. In this way we obtain
\eqn\finalAdS{
Z_{AdS}[\psi_-, \bar{\psi}_+] = \exp (-S_1) = \exp \left( - \int d^d
\xpp \int d^d \xp \bar{\psi}_+ (\xpp) \Omega(\xpp, \xp) \psi_- (\xp)
\right)\ ,
}
where
\eqn\Omegapm{
\eqalign{
\Omega (\xpp, \xp)  & = \lim_{\epsilon \rightarrow 0} \int d^d \x
\epsilon^{2m +1} \Bigl(\epsilon^2 + |\x\! -\! \xpp|^2
\Bigr)^{-{d+1\over 2}-m} \cr
& {}\qquad\qquad \times
\Bigl(\epsilon^2 + |\x\! -\! \xp|^2 \Bigr)^{-{d+1\over 2}-m} (\xpp\!
-\!
\xp) \cdot \G
\cr
& = c\ |\xpp - \xp|^{-d - 1 - 2 m} (\xpp - \xp) \cdot \G \ . \cr
}
}

\newsec{Comparison to conformal field theory}
We will now compare these results to what would be expected from a
conformal field theory on $M_d$. The expression \finalAdS\ can be
interpreted as the generating functional $Z_{CFT}[\psi_-,
\bar{\psi}_+]$ of correlation functions of a free-field operator
${\cal O}^{\alpha}$ and its conjugate $\bar{\cal O}^\alpha$ on $M_d$
with the two-point function
\eqn\twopoint{
\langle \bar{\cal O}^\beta (\xpp) {\cal O}^\alpha (\xp) \rangle =
\Omega^{\alpha \beta} (\xpp, \xp)\ ,
}
where we have written out
the spinor indices $\alpha$ and $\beta$ explicitly.

The expression \twopoint\ is in fact the correct two-point function
for a quasi-primary spinor operator of scaling dimension
\eqn\dimension{
\Delta = {d \over 2} + m \ .
}
Indeed, for an operator ${\cal O}^\alpha$ transforming in some
representation of the $O(d)$ Lorentz group and with scaling dimension
$\Delta$, conformal invariance fixes the two-point function,
up to normalization, to be \FerraraGattoGrilloParisi
\eqn\Rtwopoint{
\langle \bar{\cal O}^\beta (\xpp) {\cal O}^\alpha (\xp) \rangle
= |\xpp \! - \! \xp|^{-2 \Delta} D^{\alpha \beta} \bigl( R(\xpp,\xp)
\bigr)\ ,
}
where $D^{\alpha \beta} \bigl( R(\xpp,\xp) \bigr)$ denotes the
representation matrix in the appropriate representation of the $O(d)$
element
\eqn\Lorentz{
R^i{}_j (\xpp,\xp) = \delta^i{}_j - 2 |\xpp\! -\! \xp|^{-2}
 (x^{\prime \prime}\! - \! x^\prime)^i (x^{\prime \prime}\! -\!
x^\prime)_j\  .
}
We note that $\det(R^i{}_j)=-1$.
For the spinor representation, we have
\eqn\spinorLorentz{
D^{\alpha \beta} \bigl( R(\xpp,\xp) \bigr)
= - |\xpp\! -\! \xp|^{-1} (\xpp\! -\! \xp) \cdot \bigl(
\G \Gamma^0 \bigr)^{\alpha \beta} \ ,
}
which can be checked by verifying the invariance
of $(\Gamma^i)^{\alpha \beta}$ under the rotation \Lorentz:
$R^i{}_j D\inv \Gamma^j D = \Gamma^i$. In the subspace of definite
$\Gamma^0$
eigenvalue we are working on, we may set $\Gamma^0=-1$ in
\spinorLorentz.
Then \Rtwopoint,
with $\Delta= {d\over 2} +m$, agrees with \twopoint.

We believe that techniques similar to those used in the present
paper
are appropriate for the spin-3/2 Rarita--Schwinger field on AdS.
Having understood the AdS/CFT correspondence for free fields, a
natural
continuation would be to consider interacting theories.
It would be interesting to see how the results of this paper,
in particular the need for an extra boundary term in the action,
generalize in that context.


\listrefs

\end